
\magnification=\magstep1
\baselineskip=20 truept
\def\ni{\noindent}
\def\bn{\bigskip\noindent}
\def\sn{\smallskip\noindent}
\def\pa{\partial}

\def\half{{\textstyle{1\over2}}}
\def\vp{\varphi}
\def\ii{{\rm i}}
\nopagenumbers
\font\scap=cmcsc10

\rightline{DTP/95/3; NI94031}
\vskip 1truein
\centerline{\bf Discrete Toda Field Equations.}

\vskip 1truein
\centerline{\scap R. S. Ward}
\bn\centerline{\sl Isaac Newton Institute for Mathematical Sciences,}
\centerline{\sl Cambridge CB3 0EH, UK}
\sn\centerline{and}
\sn\centerline{\sl Dept of Mathematical Sciences, University of Durham,}
\centerline{\sl Durham DH1 3LE, UK.$^*$}

\vskip 2truein
\ni{\bf Abstract.} There are two-dimensional Toda field equations corresponding
to each (finite or affine) Lie algebra. The question addressed in
this note is whether there exist integrable discrete versions of these.
It is shown that for certain algebras (such as $A_n$, $A_n^{(1)}$
and $B_n$) there do, but some of these systems are defined on
the half-plane rather than the full two-dimensional lattice.

\bn To appear in Physics Letters A.

\vfill
\hrule
\sn$^*$ Permanent address.
\eject

\pageno=2
\footline={\hss\tenrm\folio\hss}

The original Toda chain
$$\ddot\vp_j = \exp-\sum_k C_{jk}\vp_k, \eqno(1)$$
where the $C_{jk}$ are certain constants,
is an integrable system in which $\vp_j(t)$ depends on a discrete variable
$j$ and on a continuous variable $t$ [1]. This can be generalized in two ways,
while maintaining integrability. First, the number of continuous variables
can be increased from one to two (replace $t$ by $u$,$v$),
with the second derivative being replaced
by the two-dimensional wave operator $\pa_u\pa_v$. Secondly, the discrete
variable can be made to correspond to the simple roots (Dynkin diagram) of a
finite-dimensional or affine Lie algebra. The equation now looks like
$$ \pa_u\pa_v \vp_j = \exp -\sum_k C_{jk} \vp_k, \eqno(2) $$
where $C_{jk}$ is the Cartan matrix of the Lie algebra [2--5]. In the original
(infinite) Toda chain, the Lie algebra is the $N\to\infty$ limit of
$A_{N-1}=su(N)$.

The subject of this note is the question of whether there exists an
integrable version of (2)
in which the variables $u$ and $v$ are discrete, i.e. an integrable partial
difference equation which reduces to (2) in a continuum limit.
The general topic of discrete integrable versions of partial
differential equations has long been of interest. There have been several
different approaches, for example:
\item{$\bullet$} discrete version of the AKNS linear system [6];
\item{$\bullet$} Hirota's method [7--12];
\item{$\bullet$} Zakharov-Shabat dressing method [13];
\item{$\bullet$} linear integral equations [14--17].

\ni In connection with discrete versions of the Lax system for the Toda chain
(1), see also refs 18--21. Suris [18--20] exhibited discrete integrable
versions of the Toda chain corresponding to all the non-exceptional affine Lie
algebras. The question here is whether there are analogous integrable lattice
versions of the Toda field equations (2). By \lq\lq integrable\rq\rq\ we mean
that the equation can be expressed as the consistency condition of a linear
sysytem of the type described below.
Two particular examples that have been known for some time are:
\item{$\bullet$} a discrete sine-Gordon equation [9], corresponding to
                 the algebra $A_1^{(1)}$;
\item{$\bullet$} a discrete $A_\infty$ Toda field equation [13, 12, 22].

\ni The latter is often known as the Hirota or Hirota-Miwa equation.
It is fairly straightforward to impose boundary conditions (in the
index $j$) which reduce it to a discrete $A_n$ or $A^{(1)}_n$ Toda
field equation; but the resulting finite-rank systems
do not seem to have appeared explicitly in the literature before.
Let us therefore begin by presenting these,
in a form similar to that of ref 18.

Let $u$ and $v$ be coordinates on the two-dimensional lattice ${\bf Z^2}$,
with lattice spacing $h$. So each of $u$ and $v$ takes values which are
integer multiples of $h$. A subscript denotes shift in the forward direction:
in other words, if $f(u,v)$ is a function on the lattice, then
 $ f_u(u,v) = f(u+h,v) $
and similarly for $f_v$.
Consider a linear system involving $N\times N$ matrices
$U$ and $V$ (with $N\geq2$), and a column N-vector $\psi$, of the form
$$\psi_u = U\psi, \qquad \psi_v = V\psi, \eqno(3) $$
where
$$\eqalign{U &= \lambda^{-1}I+h\exp(F_u)X_-\exp(-F), \cr
           V &= \lambda\exp(F_v-F) -hX_+. \cr} \eqno(4) $$
The various symbols appearing in (4) are defined as follows:
\item{$\bullet$} $\lambda$ is an arbitrary scalar parameter;
\item{$\bullet$} $I$ is the identity matrix;
\item{$\bullet$} $F ={\rm diag}(f_1,f_2,\ldots,f_N)$ is a diagonal matrix;
\item{$\bullet$} $X_+$ is the matrix with ones immediately above the main
     diagonal, and zeros elsewhere, i.e. $(X_+)_{ab}=\delta_{a-b+1}$;
\item{$\bullet$} $X_-$ similarly has ones just below the main diagonal.

\ni Note that $X_+$ is the sum of elements of $A_{N-1}$ corresponding to
its simple roots, in the usual representation.
We are thinking in terms of a specific representation of
the Lie algebra, because the discrete Toda equations
are not purely Lie-algebraic (unlike in the continuum case).

The consistency condition for (3) is $U_v V = V_u U$. The $\lambda^{-1}$
and $\lambda$ terms of this equation are automatically satisfied by (4), and
the remaining ($\lambda$-independent) term is equivalent to
$$ \eqalign{
  \exp(\Delta^2f_1) &= [1+h^2\exp(f_{2u}-f_{1v})], \cr
  \exp(\Delta^2f_k) &= [1+h^2\exp(f_{(k+1)u}-f_{kv})]/
                       [1+h^2\exp(f_{ku}-f_{(k-1)v})] \cr
                    &\hskip 2 truein{\rm for}\ 2\leq k \leq N-1, \cr
  \exp(\Delta^2f_N) &= 1/[1+h^2\exp(f_{Nu}-f_{(N-1)v})], \cr
} \eqno(5) $$
where
$$ \Delta^2 f = f_{uv} - f_u - f_v + f. \eqno(6) $$

The equations (5) are the discrete $A_{N-1}$ Toda field equations. Strictly
speaking, we should impose the constraint
$$ \sum_{k=1}^N f_k = 0, \eqno(7) $$
so that there are only $N-1$ independent fields. This is consistent with the
field equations, because one of the equations (5) is in effect
${\rm det}(U_v V) = {\rm det}(V_u U)$, and this is automatically satisfied
if (7) holds. So one should think of (5) as consisting of $N-1$
second-order difference equations for $N-1$ functions.

Let us examine two limiting cases of these field equations. First, if the
$f_k$ depend only on the single discrete variable $t=u+v$, then (5) reduces
to the discrete $A_{N-1}$ chain in the form described by Suris [18,19]. He does
not impose the constraint (7), but remarks that the quantity
$\Delta_t \sum f_k$ is a constant of motion. Secondly, in the continuum
limit $h\to0$ the equations (5) do indeed reduce to (2), where $C_{jk}$
is the Cartan matrix of $A_{N-1}$ and $\vp_k = \sum_{j=1}^k f_j$
for $1\leq k\leq N-1$.

The equations corresponding to the affine algebra $A_{N-1}^{(1)}$ are very
similar. The linear system is again (4), but now the constant matrix $X_+$
has an additional unit entry, in its lower left-hand corner: $(X_+)_{ab}
=\delta_{a-b+1} + \delta_{a-N} \delta_{b-1}$.
Similarly, $X_-$ has a one in its upper right-hand corner. The resulting
field equations are
$$ \exp(\Delta^2f_k) = [1+h^2\exp(f_{(k+1)u}-f_{kv})]/
                       [1+h^2\exp(f_{ku}-f_{(k-1)v})] \eqno(8) $$
for $1\leq k \leq N$, where $f_0=f_N$ and $f_{N+1}=f_1$ (in other words,
$f_k$ is periodic in the index $k$, with period $N$).

The two special cases described above for $A_{N-1}$ work in exactly the same
way for these $A_{N-1}^{(1)}$ equations. Another special case worth pointing
out is that of $A_1^{(1)}$: taking $f_1=-f_2=\ii\vp$, equation (8) with $N=2$
reduces to the discrete sine-Gordon equation for $\vp$ [9].

By making different choices of the matrices $F$ and $X_{\pm}$, Suris [18]
obtained discrete Toda chains corresponding to other affine Lie algebras.
Another way of deriving the same results is to use a symmetry reduction
(cf. ref 5 for the continuum case). For the one-dimensional chain, where
$f_k=f_k(t)=f_k(u+v)$, the equations (5) and (8) have the ${\bf Z_2}$
symmetry
$$ f_N\mapsto-f_1,\quad f_{N-1}\mapsto-f_2,
   \quad\ldots,\quad f_1\mapsto-f_N.\eqno(9) $$
So we can reduce by putting $f_N=-f_1$ etc. The $A_{N-1}$ chain then reduces
to the chain corresponding to $C_{N/2}$ (for $N$ even) or $B_{(N-1)/2}$
(for $N$ odd); and $A_{N-1}^{(1)}$ reduces to the chain $C_{N/2}^{(1)}$
(for $N$ even) or $A_{(N-1)}^{(2)}$ (for $N$ odd). Finally, the chain
$D_{N/2}^{(2)}$ is obtained from $A_{N-1}^{(1)}$ with $N$ even, via an
analogous but slightly different reduction.
Chains corresponding to other
Lie algebras such as $D_N$ involve a Lax representation of a more general type,
and do not fit into the scheme presented here.
An examination of some particular examples illustrates that these equations,
despite the notation, are not strictly Lie-algebraic. For example, the
discrete Toda chains labelled by $A_1$ and $B_1$ are different, even though
these Lie algebras are isomorphic.

For the (two-dimensional) Toda field equations (5) and (8), however, things
are not quite so simple. These equations do not admit the symmetry (9).
But they do admit such a symmetry if we also interchange the independent
variables $u$ and $v$. In other words, we can reduce (5) and (8) by
imposing
$$ f_N(u,v)=-f_1(v,u), \quad f_{N-1}(u,v)=-f_2(v,u), \ldots . \eqno(10) $$
Because of the nonlocal nature of this constraint, one then gets nonlocal
field equations. But one can restore locality by restricting to symmetric
functions of $u$ and $v$. In effect, this yields a partial difference
equation on the half-plane, rather than on the full plane ${\bf Z^2}$.
Let us examine a particular example, namely
the systems corresponding to $B_n$.

So let $N=2n+1$ be an odd integer, with $n\geq1$. Impose (10), and also the
condition that $f_{kv}$ be symmetric in $(u,v)$, ie.
$$ f_k(u,v+h) = f_k(v,u+h) \quad {\rm for}\ k=1,2,\ldots,n. \eqno(11) $$
Finally, impose $f_{n+1} = 0$, which is consistent with (5, 10, 11).
The effect of all this is that the field equations (5) reduce to
$$ \eqalign{
  \exp(\Delta^2f_1) &= [1+h^2\exp(f_{2u}-f_{1v})], \cr
  \exp(\Delta^2f_k) &= [1+h^2\exp(f_{(k+1)u}-f_{kv})]/
                       [1+h^2\exp(f_{ku}-f_{(k-1)v})] \cr
                    &\hskip 2 truein{\rm for}\ 2\leq k \leq n-1, \cr
  \exp(\Delta^2f_n) &= [1+h^2\exp(-f_{nv})]/
                       [1+h^2\exp(f_{nu}-f_{(n-1)v})], \cr
} \eqno(12) $$
together with the symmetry constraint (11).
So we can think of (12) as being defined on the half-plane $v\leq u+h$.
In terms of the laboratory coordinates $t=\half(u+v-h)$ and
$x=\half(u-v+h)$, this half-plane consists of those lattice sites which lie
in the region $x\geq0$. The field equation is (12) at lattice sites for
which $v<u+h$, and there is a boundary condition if $v=u+h$. The latter
is obtained from (12) by using the symmetry about $x=0$: for each $k$,
$f_{kv}$ is replaced by $f_{ku}$.

The continuum limit $h\to\infty$ of this gives the $B_n$ Toda field equation
in the half-plane $x\geq0$, with the boundary condition $\pa_x \vp_k=0$
on $x=0$. Notice also that the discrete $B_n$ Toda chain is a special case,
since the $f_k$ are then independent of $x$.

In conclusion, it has been shown that there exist integrable discrete versions
of the two-dimensional Toda field equations, corresponding to at least some
of the simple finite-dimensional or affine Lie algebras. The basic cases are
$A_n$ and $A_n^{(1)}$. Others such as $B_n$ and $C_n$ can be obtained from
these by reduction, but it appears that these reduced systems are defined on
the half-lattice rather than the full two-dimensional lattice. Not all the
reductions have as yet been analysed. And for the algebras $D_n$, $B_n^{(1)}$,
$D_n^{(1)}$ and $A_{2n-1}^{(2)}$ the picture is unclear: the corresponding
discrete Toda chains involve Lax representations of a more general type
[18,19], and it is an open question whether discrete two-dimensional
versions exist.
Similarly, the situation with regard to the exceptional
Lie algebras is unknown. It would be preferable to have a \lq\lq group\rq\rq\
description which did not involve a particular representation, but it is
not clear what sort of algebraic structure is required for this.

\vfill\eject

\ni{\bf References.}
\item{[1]} M. Toda, Theory of nonlinear lattices (Springer, 1988).
\item{[2]} A. V. Mikhailov, M. A. Olshanetsky and A. M. Perelomov,
           Comm. Math. Phys. 79 (1981) 473.
\item{[3]} G. Wilson, Ergod. Th. and Dynam. Syst. 1 (1981) 361.
\item{[4]} A. P. Fordy and J. Gibbons, Proc. R. Ir. Acad. A 83 (1983) 33.
\item{[5]} D. Olive and N. Turok, Nucl. Phys B 215 (1983) 470.
\item{[6]} M. J. Ablowitz and F. J. Ladik, Stud. Appl. Math. 55 (1976) 213;
           57 (1977) 1.
\item{[7]} R. Hirota, J. Phys. Soc. Jpn. 43 (1977) 1424.
\item{[8]} R. Hirota, J. Phys. Soc. Jpn. 43 (1977) 2074.
\item{[9]} R. Hirota, J. Phys. Soc. Jpn. 43 (1977) 2079.
\item{[10]} R. Hirota, J. Phys. Soc. Jpn. 45 (1978) 321.
\item{[11]} R. Hirota, J. Phys. Soc. Jpn. 46 (1979) 312.
\item{[12]} R. Hirota, J. Phys. Soc. Jpn. 50 (1981) 3785.
\item{[13]} D. Levi, L. Pilloni and P. M. Santini, J. Phys. A 14 (1981) 1567.
\item{[14]} F. W. Nijhoff, G. R. W. Quispel and H. W. Capel,
            Phys. Lett. A 97 (1983) 125.
\item{[15]} F. W. Nijhoff, H. W. Capel, G. L. Wiersma and G. R. W. Quispel,
            Phys. Lett. A 103 (1984) 293; 105 (1984) 267.
\item{[16]} G. R. W. Quispel, F. W. Nijhoff, H. W. Capel and J. van der
            Linden, Physica A 125 (1984) 344.
\item{[17]} F. W. Nijhoff, V. G. Papageorgiou and H. W. Capel, Springer
            Lecture Notes in Mathematics 1510 (1992) 312.
\item{[18]} Yu. B. Suris, Leningrad Math. J. 2 (1990) 339.
\item{[19]} Yu. B. Suris, Phys. Lett. A 145 (1990) 113.
\item{[20]} Yu. B. Suris, Phys. Lett. A 156 (1991) 467.
\item{[21]} J. Gibbons and B. A. Kupershmidt, Phys. Lett. A 165 (1992) 105.
\item{[22]} S. Saito and N. Saitoh, J. Math. Phys. 28 (1987) 1052.

\bye